\documentclass{elsart}

\usepackage{graphicx}
\usepackage{amssymb}

\begin{document}

\begin{frontmatter}

\title{Measurement of Cosmic-Ray Proton and Antiproton Spectra 
at Mountain Altitude}

%%\author[tok]{\underline{T. Sanuki}\corauthref{cor1}},
\author[tok]{T. Sanuki\corauthref{cor1}},
\corauth[cor1]{Corresponding author.}
\ead{sanuki@phys.s.u-tokyo.ac.jp}
\author[tok]{M. Fujikawa},
\author[tok]{H. Matsunaga\thanksref{tuku}},
\author[kob]{K. Abe},
\author[tok]{K. Anraku\thanksref{kanagawa}},
\author[tok]{H. Fuke},
\author[tok]{S. Haino},
\author[tok]{M. Imori},
\author[tok]{K. Izumi},
\author[kob]{T. Maeno\thanksref{cern}},
\author[kek]{Y. Makida},
\author[tok]{N. Matsui},
\author[tok]{H. Matsumoto},
\author[tok]{J. Nishimura},
\author[kob]{M. Nozaki},
\author[tok]{S. Orito\thanksref{ori}},
\author[kek]{M. Sasaki\thanksref{nasa}},
\author[kob]{Y. Shikaze},
\author[kek]{J. Suzuki},
\author[kek]{K. Tanaka},
\author[kek]{A. Yamamoto},
\author[tok]{Y. Yamamoto},
\author[kob]{K. Yamato},
\author[kek]{T. Yoshida}, and
\author[kek]{K. Yoshimura}

\address[tok]{The University of Tokyo,
 Bunkyo, Tokyo 113-0033, Japan}
\address[kob]{Kobe University,
 Kobe, Hyogo 657-8501, Japan}
\address[kek]{High Energy Accelerator Research Organization (KEK),
 Tsukuba, Ibaraki 305-0801, Japan}

\thanks[tuku]{Present address: University of Tsukuba,
 Tsukuba, Ibaraki 305-8571, Japan}
\thanks[kanagawa]{Present address: Kanagawa University,
 Yokohama, Kanagawa 221-8686, Japan}
\thanks[cern]{Present address: CERN, CH-1211 Geneva 23, Switzerland}
\thanks[ori]{deceased.}
\thanks[nasa]{Present address:
 National Aeronautics and Space Administration,
 Goddard Space Flight Center, Greenbelt, MD 20771, USA}

\begin{abstract}
Cosmic-ray proton and antiproton spectra were measured
 at mountain altitude, 2770~m above sea level. 
We observed more than $2 \times 10^5$ protons and $10^2$ antiprotons
 in a kinetic energy range between 0.25 and 3.3~GeV. 
The zenith-angle dependence of proton flux was obtained. 
The observed spectra were compared with theoretical predictions. 
\end{abstract}


\end{frontmatter}

\section{Introduction}\label{sec:intro}

Primary cosmic rays hit the Earth's atmosphere and produce
 baryons and mesons via hadronic interactions.
Absolute fluxes of these secondary cosmic rays at any altitudes
 can be calculated based on
 primary cosmic-ray intensity and interaction cross sections.
Observations of the secondary cosmic rays are very important
 for understanding the production and propagation
 of secondary cosmic-ray particles inside the atmosphere. 
It will provide information
 to verify or to improve the theoretical models. 
A number of experiments
 for measuring primary cosmic rays have been carried out
 at balloon altitude or in space,
 and atmospheric muons on the ground as well as under the ground. 
A long-period observation at mountain altitude
 with high statistics will provide a good reference
 for a study
 of secondary cosmic-ray particles inside the atmosphere.

Only a few measurements of cosmic-ray proton fluxes at mountain altitude
 were reported \cite{Kocharian54,Kocharian58,Barber80,Sembroski86}. 
Antiproton flux at mountain altitude was also reported \cite{Sembroski86}
 previous to this work. 
The reported flux was quite higher than theoretical predictions.

We report here a new observation of
 cosmic-ray protons and antiprotons at mountain altitude
 in a kinetic energy range between 0.25 and 3.3~GeV. 
A spectral shape of cosmic-ray antiprotons at mountain altitude
 was measured for the first time. 
These data will provide essentially important information
 about hadronic interactions
 between cosmic rays and atomic nuclei inside the atmosphere. 

\section{BESS Experiment}\label{sec:obs}

\subsection{Detector}
\begin{figure}
\includegraphics[width=\textwidth]{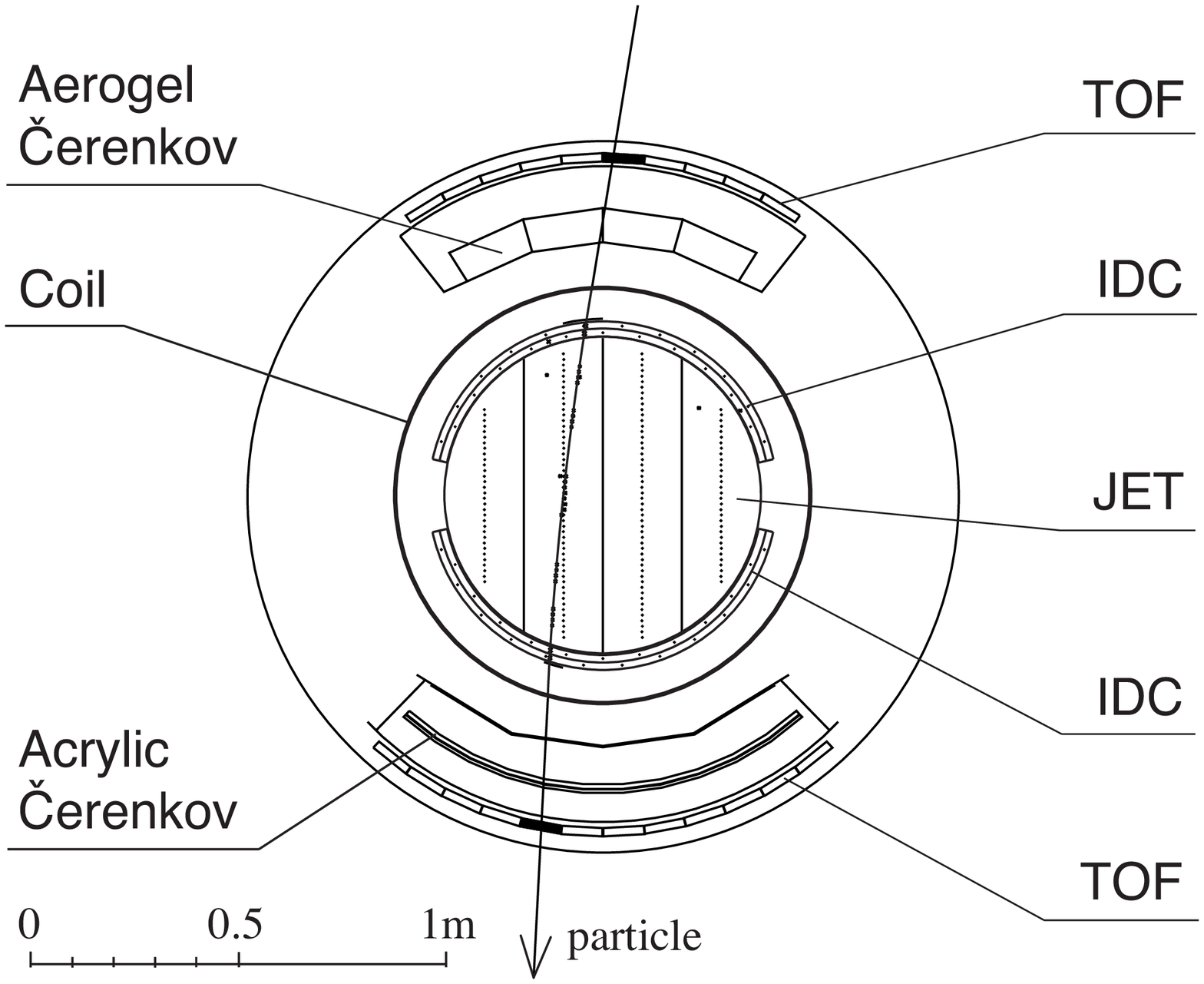}
\caption{Schematic cross-sectional view of the BESS detector.}
\label{fig:besscross}
\end{figure}

The BESS (\underline{B}alloon-borne \underline{E}xperiment
 with a \underline{S}uperconducting \underline{S}pectrometer)
 detector \cite{Orito87,Yamamoto94,agel,detector,newtof,beamtest} is
 a high-resolution spectrometer with a large acceptance
 to perform highly sensitive searches
 for rare cosmic-ray components,
 as well as precise measurement of the absolute fluxes
 of various cosmic rays. 
Fig.~\ref{fig:besscross} shows a schematic cross-sectional view
 of the BESS instrument.
In the central region,
 a uniform magnetic field of 1 Tesla is provided
 by using a thin superconducting solenoidal coil. 
A magnetic-rigidity ($R \equiv Pc/Ze$) of
 an incoming charged particle is measured
 by a tracking system,
 which consists of a jet-type drift chamber (JET)
 and two inner-drift-chambers (IDC's)
 inside the magnetic field. 
The deflection ($R^{-1}$) and its error
 are calculated
 for each event by applying a circular fitting
 using up to 28 hit points,
 each with a spatial resolution of 200~$\mu$m. 
The maximum detectable rigidity (MDR) was estimated to be 200~GV. 
Time-of-flight (TOF) hodoscopes provide the velocity ($\beta$)
 and energy loss ($\d E/\d x$) measurement. 
A 1/$\beta$ resolution of 1.6~\% was achieved in this experiment. 
In order to separate protons and antiprotons from muons,
 the BESS spectrometer is equipped with
 a threshold-type aerogel \v{C}erenkov counter. 
The refractive index of silica aerogel radiator is 1.022, 
 which corresponds to a threshold kinetic energy of 3.6~GeV
 for protons and antiprotons. 
Each particle is identified
 by requiring proper 1/$\beta$, as well as $\d E/\d x$,
 as a function of the rigidity. 
%%%%%%%%%%%%%%%%%%%%%%%%%%%%%%%%%%%%%%%%%%%%%%%%%%
%%	Fig.1					%%
%%	Schematic cross-sectional view		%%
%%%%%%%%%%%%%%%%%%%%%%%%%%%%%%%%%%%%%%%%%%%%%%%%%%

There are a plastic scintillating counter,
 an acrylic \v{C}erenkov counter and a lead plate
 just above the bottom TOF hodoscope. 
The lead plate covers about 1/5 of the total acceptance. 
These counters were used for electron/muon identification
 in the muon analysis \cite{muSanuki}.
In the analysis of protons and antiprotons discussed here,
 the signals from these counters were not examined. 

All detector components are arranged in a cylindrical configuration. 
This simple configuration and the uniform magnetic field
 result in a large geometrical acceptance and
 uniform performance in momentum measurement. 

A simple coincidence of the top- and bottom-TOF hodoscope signals
 issues the first-level ``T0-trigger.'' 
The live data-taking time is measured exactly
 by counting 1~MHz clock pulses with a scaler system
 gated by a ``ready'' status that controls the T0-trigger. 

\subsection{Observations}
The atmospheric cosmic-ray events were observed at Norikura Observatory,
 ICRR, the University of Tokyo, Japan. 
It is located at $36^\circ~06'N, 137^\circ~33'E$. 
The altitude is 2,770~m above sea level. 
The vertical geomagnetic cutoff rigidity is 11.2~GV \cite{Shea01}. 

\begin{figure}
\includegraphics[width=\textwidth]{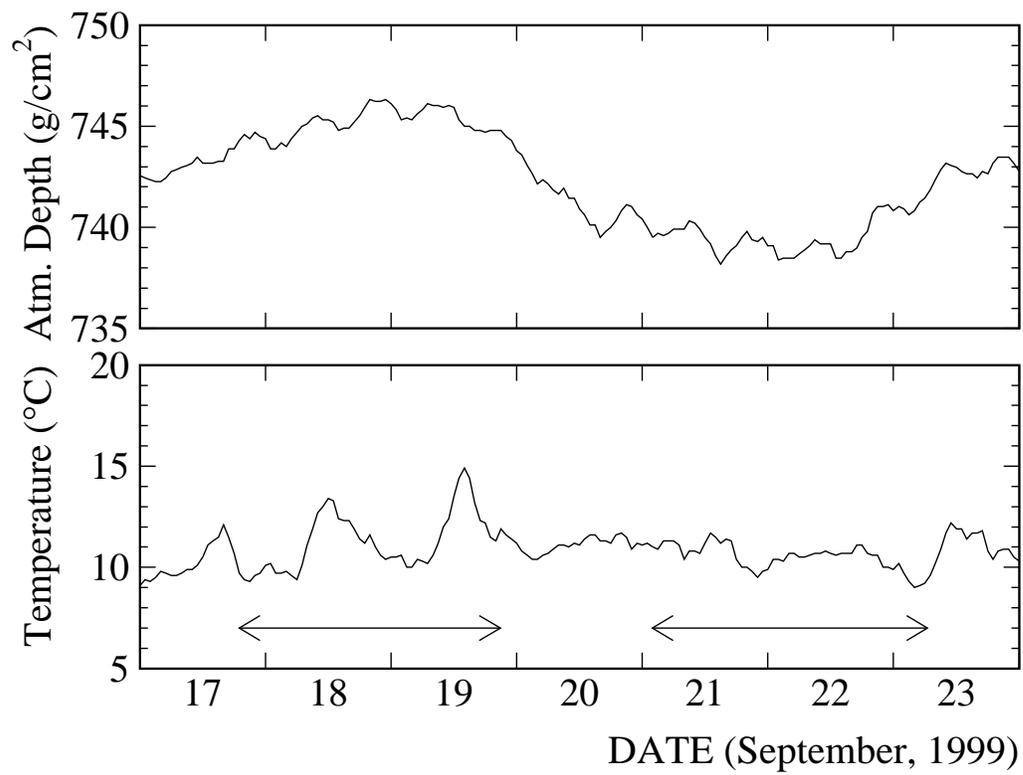}
\caption{Atmospheric depth and temperature during the observation.
 The experiment was performed
 during two periods of 17th -- 19th and 21st -- 23rd.}
\label{fig:env}
\end{figure}

The observation was performed
 during two periods of 17th -- 19th and 21st -- 23rd of September 1999. 
During the observation, the atmospheric depth and temperature
 varied as shown in Fig.~\ref{fig:env}. 
The mean (root-mean-square) atmospheric depth and temperature were
 742.4 (2.9)~${\mathrm {g/cm^2}}$ and 10.9 (1.1)~${\mathrm {^\circ C}}$,
 respectively. 
The BESS detector was operated outside the building
 and in a nylon plastic sheet tent
 so as to be less influenced by ambient conditions. 
There was no thick material within a field of view of the detector. 
A gas flow system was implemented to keep
 the purity and pressure of the gas inside the chambers
 stable enough for operation. 
%%%%%%%%%%%%%%%%%%%%%%%%%%%%%%%%%%%%%%%%%%%%%%%%%%
%%	Fig.2					%%
%%	Atmospheric depth and temperature	%%
%%%%%%%%%%%%%%%%%%%%%%%%%%%%%%%%%%%%%%%%%%%%%%%%%%

The T0-trigger rate
 was about 50~Hz. 
It is much lower than that at a usual balloon floating altitude
 of about 37~km. 
Thus, all events which satisfied the T0-trigger condition were recorded
 without imposing second-level trigger nor on-line rejections. 
This simple trigger condition reduced systematic errors. 
The live-time ratio was as high as 98.8~\% through the observation. 
The number of collected events was about $2.0 \times 10^{7}$. 

\section{Data analysis} \label{sec:analysis}

\begin{figure}
\includegraphics[width=\textwidth]{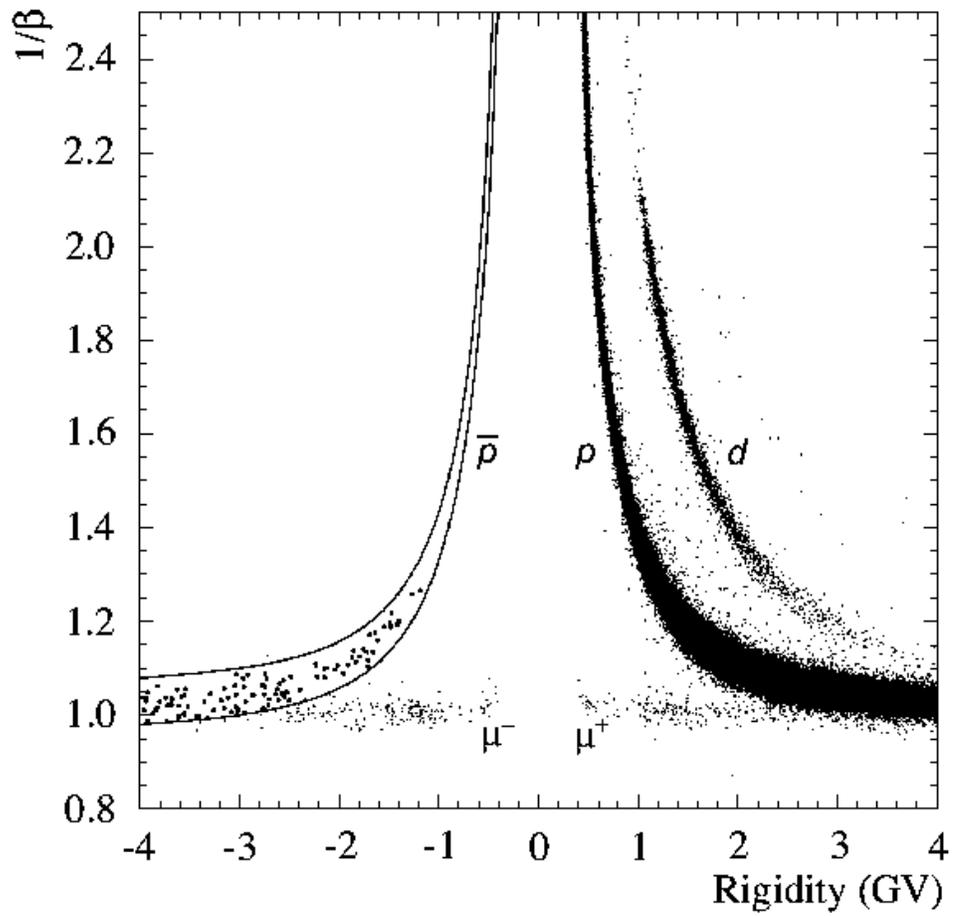}
\caption{Identification of antiproton events. 
 The solid lines define the antiproton selection band. }
\label{fig:pmbe}
\end{figure}

Particle mass was reconstructed by the velocity, rigidity and charge
 for each event, 
 and was required to be consistent with a proton and an antiproton. 
Fig.~\ref{fig:pmbe} shows $1/\beta$ distribution as a function of rigidity
 after requiring
 that there was no light output
 from the aerogel \v{C}erenkov counter,
 so as to discriminate protons and antiprotons from muons. 
%%%%%%%%%%%%%%%%%%%%%%%%%%%%%%%%%%%%%%%%%%%%%%%%%%
%%	Fig.3					%%
%%	Identification of antiproton events	%%
%%%%%%%%%%%%%%%%%%%%%%%%%%%%%%%%%%%%%%%%%%%%%%%%%%

The procedure of data analysis was almost the same as
 that of the balloon-flight data \cite{pbOrito,pbMaeno,pbAsaoka}
 except for a method of evaluating the performance of
 the aerogel \v{C}erenkov counter. 
The cosmic rays at the balloon altitude are dominated by protons. 
On the other hand,
 those at mountain altitude consist almost of muons. 
Therefore the efficiency of the aerogel \v{C}erenkov counter
 for protons and antiprotons was not obtained
 from the observed data in this analysis. 
The efficiency and its dependence on particle's velocity were estimated
 by using muons in a lower rigidity range of $|R| < 1.7$~GV,
 where muons were identified without contaminations. 
The obtained efficiency was applied to protons and antiprotons. 
This process arose an additional systematic error of 4~\%. 
In a higher rigidity range of $|R| > 1.7$~GV,
 protons and antiprotons were contaminated by muons. 
The number of contaminating muons was estimated by multiplying
 the total number of observed particles
 by the probability that a muon fakes a proton or an antiproton. 
The fake probability was evaluated to be $10^{-4}$
 by using energetic particles above 10~GeV and the same muon sample
 as was used for estimating the efficiency
 of the aerogel \v{C}erenkov counter. 
The muon background amounted to about 20~\% for the antiprotons above 2~GeV. 
It was negligibly small for the protons. 

In this analysis, the zenith angle $(\theta _z)$ was limited as
 $\cos\theta _z \geq 0.95$ for protons. 
Because of the very small flux of antiprotons, 
 the lower limit of $\cos\theta _z$ for antiproton analysis
 was relaxed to be 0.84 so as to improve its statistics. 

\section{Results and discussion}\label{result}

\begin{figure}
\includegraphics[width=\textwidth]{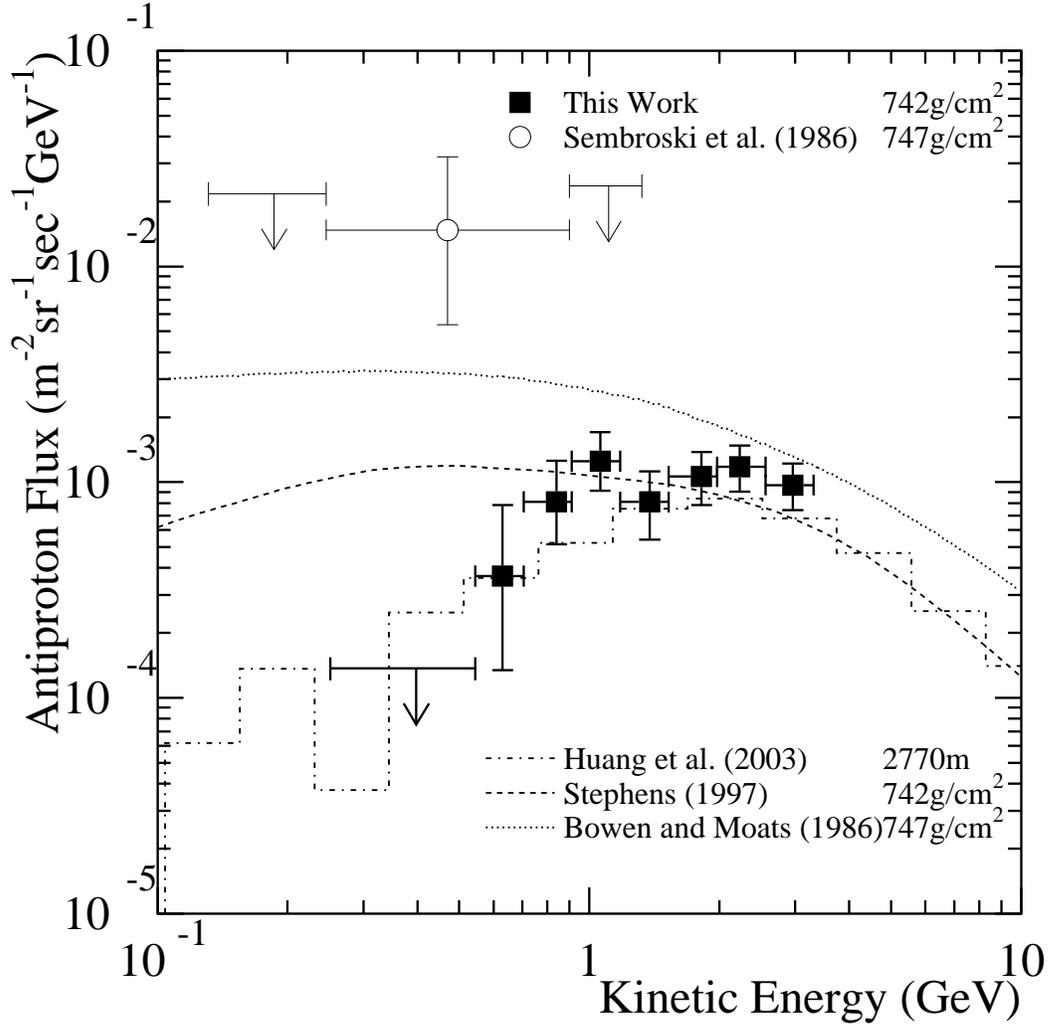}
\caption{The observed antiproton spectrum compared with the previous works,
 and with the theoretical calculations.
 The dashed line was reproduced by interpolating between two spectra
 at 700~${\mathrm {g/cm^2}}$ and 900~${\mathrm {g/cm^2}}$ using a power law.}
\label{fig:pbflux}
\end{figure}

The observed energy spectra of protons and antiprotons are shown
 in Figs. \ref{fig:pflux} and \ref{fig:pbflux}, respectively,
 together with previous measurements at mountain altitude
 \cite{Kocharian54,Kocharian58,Barber80,Sembroski86} and 
 theoretical predictions \cite{Bowen86,Stephens97,Huang03,Baret03}. 
Tables \ref{tab:pflux} and \ref{tab:pbflux} summarize
 the obtained spectra. 
The first and second errors in Tables \ref{tab:pflux} and \ref{tab:pbflux}
 represent statistical and systematic errors, respectively. 
Since the statistics of antiproton measurement was very limited,
 the statistical error was calculated as a 68.7~\% confidence interval   
 based on a ``unified approach'' \cite{Feldman98}.
%%%%%%%%%%%%%%%%%%%%%%%%%%%%%%%%%%%%%%%%%%%%%%%%%%
%%	Fig.4					%%
%%	Observed proton spectrum		%%
%%%%%%%%%%%%%%%%%%%%%%%%%%%%%%%%%%%%%%%%%%%%%%%%%%
%%%%%%%%%%%%%%%%%%%%%%%%%%%%%%%%%%%%%%%%%%%%%%%%%%
%%	Fig.5					%%
%%	Observed antiproton spectrum		%%
%%%%%%%%%%%%%%%%%%%%%%%%%%%%%%%%%%%%%%%%%%%%%%%%%%
%%%%%%%%%%%%%%%%%%%%%%%%%%%%%%%%%%%%%%%%%%%%%%%%%%
%%	Table 1					%%
%%	Observed proton spectrum		%%
%%%%%%%%%%%%%%%%%%%%%%%%%%%%%%%%%%%%%%%%%%%%%%%%%%
%%%%%%%%%%%%%%%%%%%%%%%%%%%%%%%%%%%%%%%%%%%%%%%%%%
%%	Table 2					%%
%%	Observed antiproton spectrum		%%
%%%%%%%%%%%%%%%%%%%%%%%%%%%%%%%%%%%%%%%%%%%%%%%%%%

\subsection{Protons}
\begin{figure}
\includegraphics[width=\textwidth]{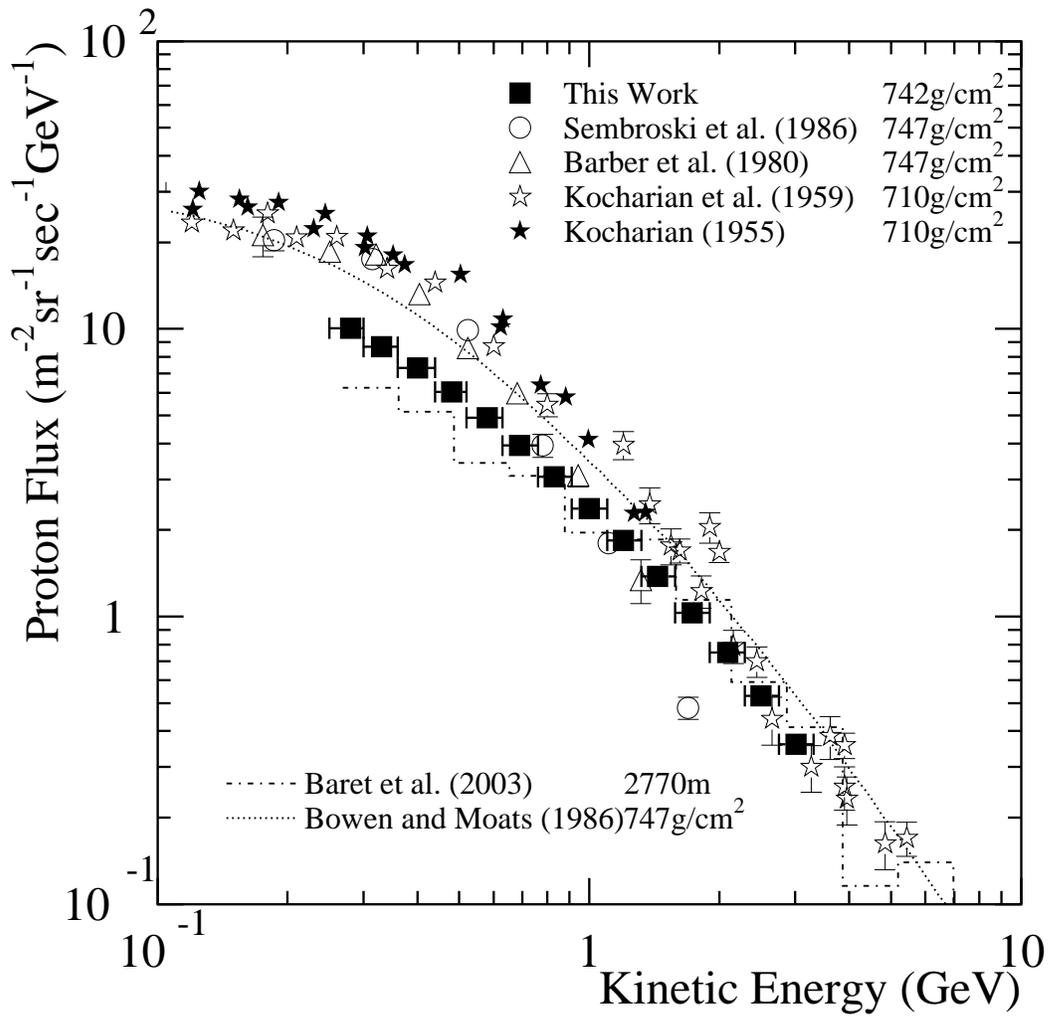}
\caption{The observed proton spectrum compared with the previous works,
 and with the theoretical calculations.}
\label{fig:pflux}
\end{figure}

There is some difference among the proton fluxes
 observed by various experiments as shown in Fig.~\ref{fig:pflux}. 
Table~\ref{table:sum_pflux_mountain} summarizes
 the locations of cosmic-ray observations at mountain altitude. 
According to our simple Monte Carlo simulations,
 the difference was generally explained
 by the different altitudes and cutoff rigidities. 
The overall spectral shape is reproduced
 by theoretical calculations \cite{Bowen86,Baret03}.
%%%%%%%%%%%%%%%%%%%%%%%%%%%%%%%%%%%%%%%%%%%%%%%%%%
%%	Table 3					%%
%%	Summary of the observations		%%
%%%%%%%%%%%%%%%%%%%%%%%%%%%%%%%%%%%%%%%%%%%%%%%%%%

\begin{figure}
\includegraphics[width=\textwidth]{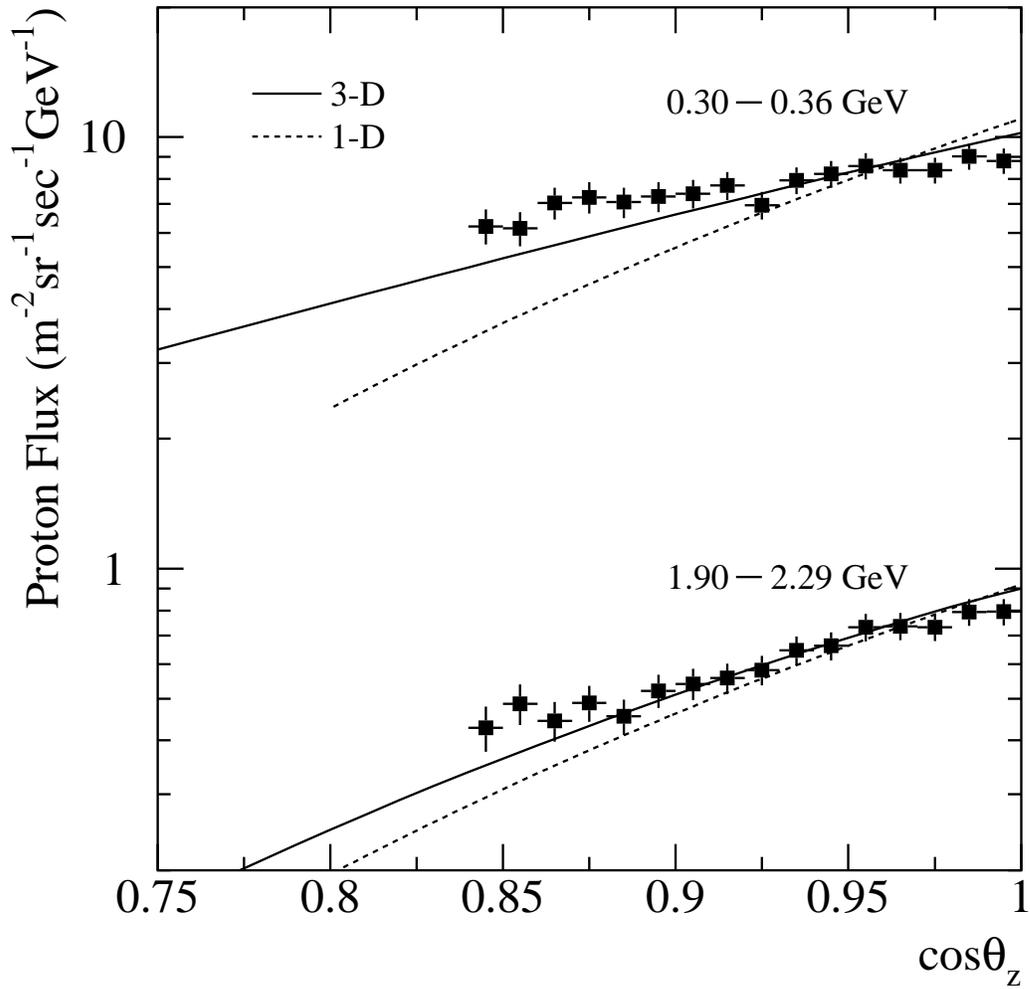}
\caption{The zenith angle dependence of proton flux. 
The dotted and solid lines show the expected dependence
in simple one-dimensional and three-dimensional calculations,
respectively.}
\label{fig:zenith-angle-proton-flux}
\end{figure}

Fig.~\ref{fig:zenith-angle-proton-flux} shows
 the zenith angle dependence of the observed proton flux, $F(\cos\theta _z)$,
 in two kinetic energy regions. 
In a simple one-dimensional approximation, 
 the zenith angle dependence can be expected as
\begin{equation}
 F(\cos\theta _z)
=F_{0}\exp
\left( \frac{X}{\Lambda} \left( 1-\frac{1}{\cos\theta _z} \right) \right)
\label{eq:zenith},
\end{equation}
 where $X$ is the atmospheric depth at the observation site
 and $\Lambda$ is the attenuation length of protons
 inside the atmosphere. 
The dashed lines in Fig.~\ref{fig:zenith-angle-proton-flux} show
 the expectation,
 in which $X/\Lambda=6$ was assumed,
 since the atmospheric depth was 742.4~${\mathrm {g/cm^2}}$
 and the proton attenuation length is about 120~${\mathrm {g/cm^2}}$. 
Normalization was done by tuning $F_{0}$ so that the total flux,
 $2\pi\int F(\cos\theta)\cos\theta d\cos\theta$,
 is equal to the observed total flux. 
Relatively good agreement is found 
 between the observed data and the expectation
 in the higher energy region. 
In the lower energy region, however,
 much larger deviation is observed in the zenith angle dependence. 
These facts are
 most likely due to the effect of angular spread of secondary protons
 produced via nuclear interactions. 
This ``three-dimensional effect'' was evaluated
 by a simple analytic method
 considering the distribution of transverse momentum
 at the production point of secondary protons.
The solid lines shown in Fig.~\ref{fig:zenith-angle-proton-flux} give
 the results of the analytic calculation,
 in which the angular spread was taken into account.
They reproduce the observed data in the whole energy range
 better than the expectation by the one-dimensional calculation. 
%%%%%%%%%%%%%%%%%%%%%%%%%%%%%%%%%%%%%%%%%%%%%%%%%%
%%	Fig.6					%%
%%	Zenith angle dependence of proton flux	%%
%%%%%%%%%%%%%%%%%%%%%%%%%%%%%%%%%%%%%%%%%%%%%%%%%%

As mentioned above,
 the proton flux shown in Fig.~\ref{fig:pflux} was obtained
 by using the events satisfying the zenith angle condition of
 $\cos\theta _z \geq 0.95$. 
The average zenith angle, $\langle\cos\theta _z\rangle$, was 0.98. 
The vertical flux $F_0 \equiv F(\cos\theta _z \rightarrow 1)$ was obtained
 by fitting the observed fluxes
 using a formula (\ref{eq:zenith}),
 where ``$X/\Lambda$'' was used as a fitting parameter.
The $F_{0}$ obtained by fitting the data in $\cos\theta _z\ge0.95$ is
 summarized in the last column of Table~\ref{tab:pflux}. 
The systematic error due to this fitting procedure was estimated
 to be around 5~\%
 by checking variations of $F_{0}$ in changing the fitting regions. 

\subsection{Antiprotons}
Fig.~\ref{fig:pbflux} shows that
 the obtained antiproton spectrum above 1~GeV is generally consistent
 with theoretical predictions obtained
 through transport equation calculations
 by Bowen and Moats \cite{Bowen86} and Stephens \cite{Stephens97},
 as well as through a three-dimensional Monte Carlo simulation
 by Huang et~al. \cite{Huang03}. 
In the lower energy region, however,
 the transport equation calculations
 show significant disagreement with our results. 
Those three calculations predict similar antiproton spectrum above 0.3~GeV
 at a thin residual atmosphere of 5~${\mathrm {g/cm^2}}$,
 but their predictions do not agree so well with each other
 at mountain altitude \cite{Bowen86,Stephens97,Huang03}. 
It means that the shapes of antiproton production spectra
 are very similar to each other,
 and that the difference
 at mountain altitude
 is to be made
 mainly due to the different treatment of the propagation
 inside the atmosphere. 
The production spectrum of antiprotons has a sharp peak
 around 2~GeV \cite{Stephens97}. 
The flux at 0.5~GeV is as small as 1/10 of the peak flux.
Therefore most antiprotons observed below 1~GeV are
 tertiary antiprotons, 
 those which have been produced inside the atmosphere
 and then lost their energy
 during the propagation inside the atmosphere. 
In this case, 
 interaction processes of $\bar{p} + A$ (nuclei) have to be precisely treated
 for an accurate evaluation of antiproton spectrum at mountain altitude. 
In the calculation of Refs.~\cite{Bowen86,Stephens97},
 the probability that an antiproton with initial energy of $E_0$
 possesses energy $E$ after a collision was assumed
 to be uniform from $E=0$ to $E=0.90 E_0$ or $0.95 E_0$. 
On this assumption,
 the average energy after a collision is about a half of the initial energy. 
In the calculation of Ref.~\cite{Huang03}, on the other hand,
 only annihilation channels were taken into account
 in the inelastic interactions. 
Non-annihilating inelastic processes were not included
 as a process of antiproton energy loss, thus
 a tertiary antiproton remains as energetic as before the interaction. 
Our result suggests that 
 the non-annihilation process does not make significant contribution
 to the inelastic interactions of antiprotons
 during their propagation inside the atmosphere. 
Thus, the antiprotons are considered not to lose their energy considerably
 in the non-annihilating inelastic processes. 

The antiproton flux measured by Sembroski et~al. \cite{Sembroski86}
 is much higher than our measurements
 in spite of the similar observational conditions
 such as altitude (Table~\ref{table:sum_pflux_mountain}).
In that experiment, a particle trajectory was measured
 with six chambers only outside the magnetic field. 
No information about the particle trajectory was obtained
 inside the magnetic field. 
It might lead to some difficulty in strict identification of antiprotons. 

\section{Summary}\label{conclusion}

We have measured proton and antiproton spectra
 in a kinetic energy range of 0.25 -- 3.5~GeV
 at Mt. Norikura, 2,770~m above sea level, in Japan. 
The vertical geomagnetic cutoff rigidity is 11.2~GV. 
The mean atmospheric depth during the measurement was
 742.4~${\mathrm {g/cm^2}}$.

The measured proton spectrum is consistent with the previous results,
 if the different altitudes and cutoff rigidities at their observation sites
 are taken into account. 
The zenith angle dependence of the proton flux was observed. 
It suggests importance of a three-dimensional effect
 of the angular spread in secondary baryon productions
 especially in a low energy region. 

The measured antiproton flux above 1~GeV generally agrees
 with the theoretical predictions. 
In the lower energy region, however,
 our flux shows much better agreement with the calculation
 performed in assuming
 that tertiary antiprotons remain as energetic as before interactions. 
Our result suggests that 
 the energy loss of antiprotons due to non-annihilation process
 is not significant
 during their propagation inside the atmosphere.

\begin{ack}
This study was supported by a Joint Research Program of
 ICRR, the University of Tokyo. 
We would like to thank all staffs at the Norikura Observatory
 for their cooperation and helpful suggestions. 
We are indebted to Professor M.~Bu\'{e}nerd and Professor L.~Derome of
 Laboratoire de Physique Subatomique et de Cosmologie, IN2P3/CNRS and
 Dr. C.~Y.~Huang of Max-Planck-Institut f\"{u}r Kernphysik
 for helpful discussions on this issue.
We would like to thank KEK and ICEPP, the University of Tokyo
 for their continuous support and encouragement during this study. 
This experiment was supported by
 Grants-in-Aid, KAKENHI(11694104, 11440085, 09304033), from
 Ministry of Education, Culture, Sports, Science and Technology, MEXT, and
 Japan Society for the Promotion of Science, JSPS, in Japan. 
\end{ack}

\clearpage
%
% References
%

\clearpage
%
% Figures
%

\clearpage

\clearpage

\clearpage

\clearpage

\clearpage

\clearpage
%
% Tables
%
\begin{table}
\caption{Observed spectrum of protons.}
\label{tab:pflux}
\begin{tabular}{cccclrr}
\hline
\multicolumn{2}{l}{Kinetic Energy} &
\multicolumn{1}{c}{Number of $p$'s} & $\langle\cos \theta _z\rangle$ &
&\multicolumn{2}{c}{Proton Flux} \\
\cline{6-7}
Range & Mean & & && \multicolumn{1}{c}{$\cos\theta _z \ge 0.95$} & \multicolumn{1}{c}{$\cos\theta _z \rightarrow 1$} \\
(GeV) & (GeV) & & && \multicolumn{2}{c}{(${\mathrm {m^{-2}sr^{-1}sec^{-1}GeV^{-1}}}$)} \\
\hline
0.25 -- 0.30 & 0.28 & 10528 & 0.98 &&10.05 $\pm$ 0.10 $\pm$ 0.46 & 10.70 $\pm$ 0.10 $\pm$ 0.73 \\
0.30 -- 0.36 & 0.33 & 11472 & 0.98 && 8.66 $\pm$ 0.08 $\pm$ 0.40 &  8.89 $\pm$ 0.08 $\pm$ 0.61 \\
0.36 -- 0.44 & 0.40 & 12048 & 0.98 && 7.33 $\pm$ 0.07 $\pm$ 0.34 &  7.81 $\pm$ 0.07 $\pm$ 0.53 \\
0.44 -- 0.52 & 0.48 & 12219 & 0.98 && 6.04 $\pm$ 0.05 $\pm$ 0.28 &  6.26 $\pm$ 0.06 $\pm$ 0.43 \\
0.52 -- 0.63 & 0.58 & 12014 & 0.98 && 4.91 $\pm$ 0.04 $\pm$ 0.23 &  5.16 $\pm$ 0.05 $\pm$ 0.35 \\
0.63 -- 0.76 & 0.69 & 11487 & 0.98 && 3.93 $\pm$ 0.04 $\pm$ 0.18 &  4.17 $\pm$ 0.04 $\pm$ 0.28 \\
0.76 -- 0.91 & 0.83 & 10711 & 0.98 && 3.06 $\pm$ 0.03 $\pm$ 0.14 &  3.21 $\pm$ 0.03 $\pm$ 0.22 \\
0.91 -- 1.10 & 1.00 & ~9664 & 0.98 && 2.37 $\pm$ 0.02 $\pm$ 0.12 &  2.52 $\pm$ 0.03 $\pm$ 0.18 \\
1.10 -- 1.32 & 1.20 & ~8889 & 0.98 && 1.84 $\pm$ 0.02 $\pm$ 0.09 &  1.94 $\pm$ 0.02 $\pm$ 0.14 \\
1.32 -- 1.58 & 1.44 & ~7803 & 0.98 && 1.38 $\pm$ 0.02 $\pm$ 0.07 &  1.47 $\pm$ 0.02 $\pm$ 0.11 \\
1.58 -- 1.90 & 1.73 & ~6877 & 0.98 && 1.03 $\pm$ 0.01 $\pm$ 0.05 &  1.07 $\pm$ 0.01 $\pm$ 0.08 \\
1.90 -- 2.29 & 2.09 & ~5947 & 0.98 && 0.75 $\pm$ 0.01 $\pm$ 0.04 &  0.80 $\pm$ 0.01 $\pm$ 0.06 \\
2.29 -- 2.75 & 2.50 & ~4940 & 0.98 && 0.53 $\pm$ 0.01 $\pm$ 0.03 &  0.57 $\pm$ 0.01 $\pm$ 0.04 \\
2.75 -- 3.31 & 3.01 & ~4026 & 0.98 && 0.36 $\pm$ 0.01 $\pm$ 0.02 &  0.39 $\pm$ 0.01 $\pm$ 0.03 \\
\hline
\end{tabular}
\end{table}

\clearpage

\begin{table}
\caption{Observed spectrum of antiprotons.}
\label{tab:pbflux}
\begin{tabular}{ccccclr}
\hline
\multicolumn{2}{l}{Kinetic Energy} &
\multicolumn{1}{c}{Number of $\bar{p}$'s} & \multicolumn{1}{c}{Number of BG's} & $\langle\cos\theta _z\rangle$ &
&\multicolumn{1}{c}{Antiproton Flux} \\
\cline{7-7}
Range & Mean & & & &&\multicolumn{1}{c}{$\cos\theta _z \ge 0.84$} \\
(GeV) & (GeV) & & & && \multicolumn{1}{c}{(${\mathrm {m^{-2}sr^{-1}sec^{-1}GeV^{-1}}}$)} \\
\hline
0.25 -- 0.54 &  --  & ~0 & 0.0 &  --  && 1.37$                          \times 10^{-4}$ upper limit \\
0.54 -- 0.70 & 0.63 & ~2 & 0.0 & 0.97 && 3.67$^{+4.14+0.17}_{-2.32-0.17}\times 10^{-4}$ \\
0.70 -- 0.91 & 0.84 & ~6 & 0.0 & 0.97 && 8.11$^{+4.44+0.38}_{-2.94-0.38}\times 10^{-4}$ \\
0.91 -- 1.18 & 1.06 & 12 & 0.0 & 0.94 && 1.25$^{+0.45+0.08}_{-0.33-0.08}\times 10^{-3}$ \\
1.18 -- 1.53 & 1.38 & 10 & 0.1 & 0.96 && 8.09$^{+3.10+0.53}_{-2.62-0.53}\times 10^{-4}$ \\
1.53 -- 1.98 & 1.82 & 18 & 1.3 & 0.96 && 1.06$^{+0.31+0.07}_{-0.27-0.07}\times 10^{-3}$ \\
1.98 -- 2.56 & 2.23 & 29 & 5.2 & 0.96 && 1.18$^{+0.29+0.08}_{-0.26-0.08}\times 10^{-3}$ \\
2.56 -- 3.31 & 2.96 & 33 & 7.8 & 0.94 && 9.70$^{+2.43+0.63}_{-2.18-0.63}\times 10^{-4}$ \\
\hline
\end{tabular}
\end{table}

\clearpage

\begin{table}
\caption{\label{table:sum_pflux_mountain}
Summary of the observation sites at mountain altitude.}
\begin{tabular}{lccc}
\hline
                  & This Work      & Refs. \cite{Kocharian54,Kocharian58}& Refs. \cite{Barber80,Sembroski86}\\
\hline
Site              & Mt. Norikura   & Mt. Aragats    & Mt. Lemmon     \\
Altitude          & 2770~m         & 3200~m         & 2750~m         \\
Atmospheric Depth & 742~${\mathrm {g/cm^2}}$ & 710~${\mathrm {g/cm^2}}$ & 747~${\mathrm {g/cm^2}}$ \\
Cutoff Rigidity   & 11.2~GV        & 7.6~GV         & 5.6~GV         \\
\hline
\end{tabular}
\end{table}

\end{document}